\documentclass[preprint,12pt]{elsarticle}
\usepackage{amssymb}
\usepackage{hyperref} 
\usepackage{graphicx}    
\usepackage{natbib}        
\usepackage{algorithm}
\usepackage{algorithmicx}
\usepackage{algpseudocode}
\usepackage{amsmath,amssymb,amsfonts}
\usepackage{textcomp}
\usepackage{xcolor}
\usepackage{subcaption}
\usepackage{mathtools, cuted}

\usepackage{tikz}
\usetikzlibrary{shapes.geometric, arrows, positioning}
\definecolor{lightgreenishblue}{RGB}{200,220,250}
\tikzstyle{startstop} = [rectangle, rounded corners, minimum width=4.1cm, minimum height=1.5cm, align = left, draw=black, fill=lightgreenishblue, text width=4.1cm]
\tikzstyle{process} = [rectangle, rounded corners, minimum width=4.1cm, minimum height=1.5cm, align = left, draw=black, fill=lightgreenishblue, text width=4.1cm]
\tikzstyle{arrow} = [ultra thick,->,>=stealth]

\date{}
\begin{document}

\begin{frontmatter}

\title{Some contributions to Lagrangian modelling of Power Converters}

\author[1,2]{Mosaib Ul Munieeb}
\author[1,3]{Shakir Showkat Sofi \corref{cor1}} \ead{shakir.sofi@skoltech.ru}
\author[1]{Fazil Bashir}
\author[1]{Munieeb Ul Hassan}
\author[1]{Shahkar Ahmad Nahvi} 
            
\affiliation[1]{organization={Dept. Electrical Engineering, IUST},
            city={Awantipora},
            state = {J\&K},
            postcode={192122}, 
            country={India}}
\affiliation[2]{organization={Dept. Electrical Engineering, NTUST},
            city={Taipei},
            postcode={10607}, 
            country={Taiwan}}

\affiliation[3]{organization={Dept. Electrical Engineering (ESAT), KU Leuven},
            city={Leuven},
            postcode={3000}, 
            country={Belgium}}
\fntext[fn1]{At the time of the research, all the authors were affiliated to the Department of Electrical Engineering, IUST, J\&K, 192122, India.}

\begin{abstract}
Lagrangian modelling can be used to derive mathematical models for complex power electronic converters. This approach uses scalar quantities (kinetic and potential energy) to derive models, which is simpler than using (vector-based) force balance equations. It employs generalized coordinates, making it easier to deal with complex systems with constraints. This systematic approach results in equations that can be expressed in state-space form, which allows for the simplification of the simulation and design process and the use of many standard software packages for system analysis and simulation. In this work, contributions are made regarding the procedure to be followed for the Lagrangian modelling of power converters and the incorporation of constraints within the Lagrangian framework. Furthermore, for the first time, Lagrangian modelling is extended to non-ideal, high-fidelity descriptions of standard power electronic circuits.
\end{abstract}

\begin{keyword}
Modelling \sep Euler-Lagrangian models \sep energy \sep power converters \sep switched systems.
\end{keyword}

\end{frontmatter}

\section{Introduction}
Power electronic converters find increasing applications in devices and systems like computers, cell phones, domestic appliances, cars, aeroplanes, industrial processes, medical applications, communication systems, transportation systems and high-power electrical transmission systems. Modelling and simulation of power electronic converters are required for verification, testing and optimisation of their design. It is required to conceptualise and fabricate power electronic converters in stages \citep{mohan}, starting with an ideal circuit and gradually incorporating complex phenomena like parasitics. Accurately
simulating power electronic converters is required in many applications, e.g., ones which require their dynamic characterisation or evaluation of phenomena like electromagnetic interference. This is especially true for power electronic converters that process power in the range of megawatts, wherein the requirement for accurate simulation is more stringent and includes accounting for distributed stray parameters to obtain the energy balance of various transient topological models\cite{tan}.

Energy and its conversion is a basic, underlying phenomenon within all physical systems. Energy is a scalar quantity, and obtaining important system information from it is easier and more systematic in general, for example, in complex mechanical systems, using Energy-based methods for deriving models is easier than using force balance equations. Modelling of mechanical systems based on energy was introduced by Euler and Lagrange in the 1750s. Analogies between mechanical and electrical systems are ubiquitous \cite{bloch} and are used to derive equations of motion for mechanical systems using electrical models. Representing physical properties in terms of energy and power for power electronic converters helps in obtaining a deeper insight into the workings of the converters. Examples of such methods include models based on Brayton-Moser equations \citep{pbm}, port-Hamiltonian \citep{ortega_automatica}
and Euler-Lagrangian (EL) models \cite{Cuk,scherpen_2003, scherpen_2019, umetani2016, russer2012}. One advantage of representing the physical properties of power electronic converters in terms of energy and power is that these properties can be used for controller design. For example, the EL modelling framework causes the dynamical equations to clearly reflect energy storage, which is required in the design of passivity-based controllers \citep{ortega_book}.


In this paper, we make three distinct contributions. First, we revisit the step-by-step procedure for EL modelling of switching circuits discussed in detail in \cite{scherpen_2003, umetani2016, scherpen_99, yildiz2009} and point out cases where it is not directly applicable or gives erroneous results. We also suggest modifications to the procedure in order to improve it. Secondly, we discuss in detail the issue of incorporation of constraints in EL modelling. We deliberate on the contention \citep{scherpen_2003} that the choice of canonical coordinates charge and current, and the corresponding Lagrangian mean that Kirchoff current law is not included in the framework and point out the fact that proper labelling of system variables (in this case, in terms of the currents flowing through the dynamic circuit elements) lead to automatic incorporation of constraints in the framework and EL modelling in the unconstrained form. Lastly, we describe for the first time how to derive switched-state space models in the standard form of high-fidelity descriptions of power converters using the EL formulation.

The outline of the paper is as follows. In Section-II, the basic formulation of the EL method is reviewed, and its application to switching circuits is revisited. In Section-III, the issue of incorporation of constraints in the EL formulation is discussed, and it is demonstrated that they can be automatically included if proper labelling is done for currents flowing in the circuit. Section IV discusses high-fidelity equivalent circuits of power electronic converters and demonstrates the application of the EL method to these high-fidelity models.
\section{The EL formulation procedure and application to switching networks}
\label{inad}
The EL equation is a second-order non-linear partial differential equation written in terms of generalised coordinates $z \in \mathbb{R}^n$ and given by:
\begin{equation} \label{Eq1}
\frac{d}{d t} \Biggl(\frac{\partial \mathcal{L}(z,\dot{z})}{\partial \dot{z}}\Biggr)-\frac{ \partial \mathcal{L}(z,\dot{z})}{\partial z}=-\frac{ \partial \mathcal{D}(\dot{z})}{\partial \dot{z}}+\mathcal{F}(z),
\end{equation}
where $\mathcal{L}(z,\dot{z})=\mathcal{T}(z,\dot{z})-\mathcal{V}(z)$ is the Lagrangian which is the difference between kinetic energy $\mathcal{T}(z,\dot{z})$ and potential energy $\mathcal{V}(z)$ of the system, $\mathcal{D}(\dot{z})$ is the Rayleigh dissipation constant and $\mathcal{F} \in \mathbb{R}^n$ is the set of generalised forcing functions associated with each generalised coordinate.

The equation (\ref{Eq1}) is generally used for non-conservative systems, for conservative systems, the dissipation term is not present. The application of this method to electrical systems can be done in two ways \citep{jmeisel}. In the \textit{loop formulation}, the state variables are electric charge $q$ and current $\dot{q}$, the Lagrangian is the difference between magnetic co-energy and electrical field energy, the forcing functions being voltages in the circuit loops. In the \textit{nodal formulation}, the state variables are the magnetic flux $\varphi$ and voltage $\dot{\varphi}$, the Lagrangian is the difference between magnetic energy and electrical co-energy, the forcing function being currents at the nodes.

A comprehensive methodology for EL modelling of switching electrical networks (like the ones found in Power electronics) is enumerated in \citep{scherpen_2003}. This methodology uses the \textit{loop formulation}, hence charge $q$ and current $\dot{q}$ are taken as the dynamic variables, and the following constraint form of the EL equations is written:
\begin{eqnarray}\label{elc}
\begin{split}
&\frac{d}{d t} \Biggl( \frac{ \partial \mathcal{L}(q,\dot{q})}{\partial \dot{q}}\Biggr)-\frac{ \partial \mathcal{L}(q,\dot{q})}{\partial q}=\\
& -\frac{\partial \mathcal{D}(\dot{q},u)}{\partial \dot{q}}+A(q,u)\lambda+\mathcal{F}(q,u);
\end{split}\\
A(q,u)^T\dot{q}=0.\label{eld}
\label{constr}
\end{eqnarray}
Here $\mathcal{D}(\dot{q},u)$, $\mathcal{F}(q,u)$ -  the Rayleigh dissipation function and the generalised forcing function are functions of the switch position $u$, the constraint equations obtained from Kirchoff's current laws are given by (\ref{constr}) and $\lambda$ is the Lagrange multiplier. A discussion on why the constraint form of the EL equation is required is given in \cite{scherpen_2003}, which basically says that the unconstrained EL equations constitute a voltage balance and since Kirchoff's current law (required for circuits with more than one mesh) is not included in the framework, it has to be incorporated using the constraint form.  \par

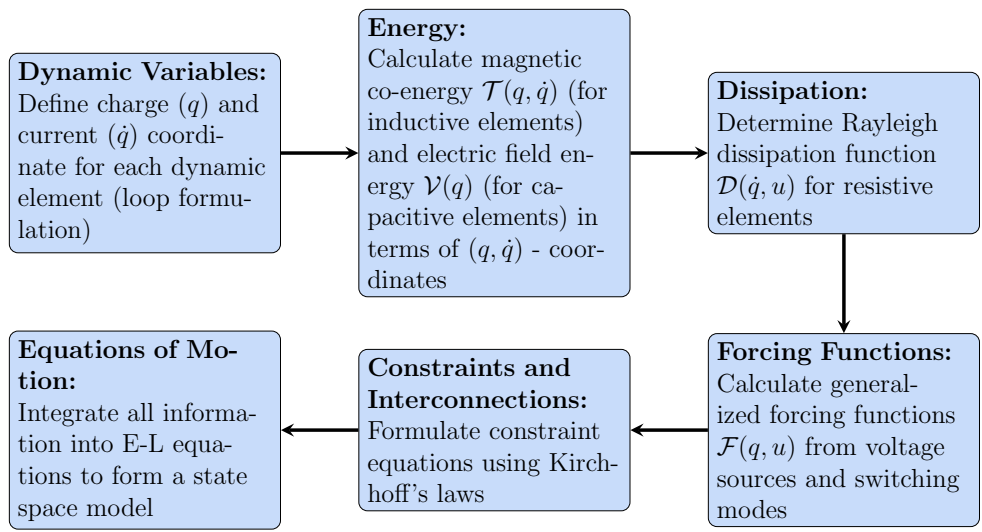
\begin{figure}[h!]
\centering
\resizebox{0.95\textwidth}{!}{%
\begin{tikzpicture}[node distance=1.25cm]
\node (start) [startstop] {\textbf{Dynamic Variables:}\\ Define charge $(q)$ and current $(\dot{q})$ coordinate for each dynamic element (loop formulation)};
\node (energy) [process, right=of start] {\textbf{Energy:}\\ Calculate magnetic co-energy $\mathcal{T}(q, \dot{q})$ (for inductive elements) and electric field energy $\mathcal{V}(q)$ (for capacitive elements) in terms of $(q, \dot{q})$ - coordinates};
\node (dissipation) [process, right=of energy] {\textbf{Dissipation:}\\ Determine Rayleigh dissipation function $\mathcal{D}(\dot{q},u)$ for resistive elements};
\node (motion) [process, below=of start] {\textbf{Equations of Motion:}\\ Integrate all information into E-L equations to form a state space model};
\node (constraints) [process, right=of motion] {\textbf{Constraints and Interconnections:}\\ Formulate constraint equations using Kirchhoff’s laws};
\node (forcing) [process, right=of constraints] {\textbf{Forcing Functions:}\\ Calculate generalized forcing functions $\mathcal{F}(q, u)$ from voltage sources and switching modes};

\draw [arrow] (start) -- (energy);
\draw [arrow] (energy) -- (dissipation);
\draw [arrow] (dissipation) -- ++(0,-1.85) -| (forcing);
\draw [arrow] (forcing) -- (constraints);
\draw [arrow] (constraints) -- (motion);
\end{tikzpicture}
}
\caption{Procedure for obtaining state space representation of switched electrical circuits using EL formulation}
\label{fig:ELprocedure}
\end{figure}

Before we proceed further, the EL procedure based on loop formulation is outlined in Figure~\ref{fig:ELprocedure} (for details, see, e.g., \cite{scherpen_2003}).

In this section, we demonstrate that the previous methodology (especially the one given in \cite{scherpen_2003}) for applying the EL formulation to switched circuits can give erroneous results. We do this using a simple example and also point out the reasons for the error and the means to correct it.
\begin{figure}[h!]
\vspace{-10mm}
\centering
\includegraphics[width=0.65\linewidth]{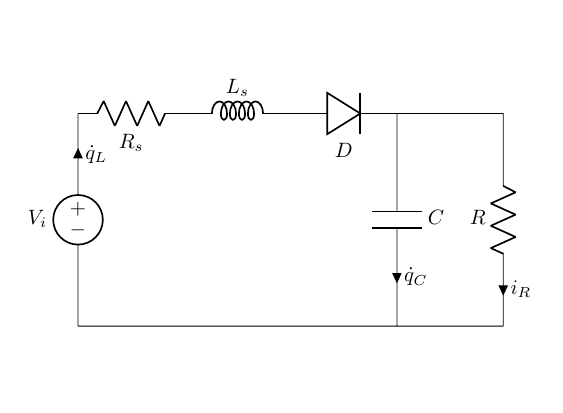}
\vspace{-10mm}
\caption{Simple diode circuit}  \label{fig0}
\end{figure}


Consider the diode circuit shown in the Fig. \ref{fig0}. Using the methodology outlined in \cite{scherpen_2003}, we select the generalised coordinates as $q_L,q_C,\dot{q}_L,\dot{q}_C$ and write the following expressions for the magnetic co-energy, the electric field energy and the forcing function vector for the circuit:
\begin{eqnarray}
\mathcal{T}(q,\dot{q})=\frac{1}{2}L_s\dot{q_L}^2;\mathcal{V}(q)=\frac{{q_C}^2}{2C};\\
\mathcal{F}(q,u) =\left[\begin{array}{cc}
uV_i & 0 
\end{array}\right]^T.
\label{arr}
\end{eqnarray}
Consequently the expression for the Lagrangian $\mathcal{L}(q,\dot{q})=\mathcal{T}(q,\dot{q})-\mathcal{V}(q)$ and the Rayleigh dissipation function can be written as:
\begin{eqnarray}\label{eel}
\mathcal{L}(q,\dot{q})=\frac{1}{2}L_s\dot{q_L}^2-\frac{{{q}_C}^2}{2C},\\ 
\mathcal{D}(\dot{q},u)=\frac{1}{2}(\dot{q_L}^2R_s+(u\dot{q}_L-\dot{q}_C)^2R).
\label{diss}
\end{eqnarray}
It is important to note that the switching function $u$ takes a value equal to either $1$ or $0$ depending on whether the diode is conducting or not. This determines whether the source $V_i$ is connected to the circuit or not and consequently, gives the expression (\ref{arr}) for ${F}(q,u)$. Further, the current through the resistor $i_R$ is either $\dot{q}_L-\dot{q}_C$ for $u=1$ or $-\dot{q}_C$ for $u=0$, and hence the expression for $\mathcal{D}(\dot{q},u)$ in (\ref{diss}). The equation (\ref{diss}) also takes care of the constraint equation obtained from Kirchoff's current law by relating $i_R,\dot{q}_L$ and $\dot{q}_C$, so $A(\dot{q},u)=0$.

Substituting (\ref{eel},\ref{diss}) in (\ref{elc},\ref{eld}) gives:
\begin{eqnarray} \label{x1}
L_s\ddot{q}_L=-(R_s+u^2R)\dot{q}_L+uR\dot{q}_C+uV_i,\\ 
\frac{q_C}{C}=(u\dot{q}_L-\dot{q}_C)R. \label{xz}
\end{eqnarray}
Substituting for $\dot{q}_C$ from (\ref{xz}) in (\ref{x1}) and rearranging gives the following switched state space model for the circuit:  
\begin{eqnarray}\label{xx}
\left[\begin{array}{c}
\ddot{q_L}\\ \\
\frac{\dot{q}_C}{C}
\end{array}\right]=\left[\begin{array}{cc}-\frac{R_s}{L_s}&\frac{-u}{L_s}\\ \\\frac{u}{C}&\frac{-1}{RC}\end{array}\right]\left[\begin{array}{c}
\dot{q_L}\\ \\
\frac{q_C}{C}\end{array}\right]+\left[\begin{array}{c}
\frac{u}{L_s}\\ \\
0
\end{array}\right]V_i.
\end{eqnarray}
The model obtained in (\ref{xx}) is a correct representation of the system dynamics for $u=1$ but does not give the correct set of equations for $u=0$.

After a closer inspection it becomes clear that the expressions (\ref{eel},\ref{diss}) written following the procedure outlined in \cite{scherpen_2003} are not correct. The insistence in \cite{scherpen_2003} that $\mathcal{L}$ should not be a function of the switch position $u$ is not right. To investigate further, we write separate expressions for the quantities $\mathcal{L}(q,\dot{q})$, $\mathcal{D}(\dot{q},u)$, $\mathcal{F}(q,u)$ and $A(\dot{q},u)$ for the two modes of operation of the circuit.

For $u=1$ the expressions are: 
\begin{eqnarray}
\mathcal{T}(q,\dot{q}) = \frac{1}{2}L_s\dot{q_L}^2 ; \mathcal{V}(q) = \frac{{q_C}^2}{2C} ; \mathcal{F}(q,u) = \left[\begin{array}{cc}
V_i & 0 
\end{array}\right]^T;\\
\mathcal{L}(q,\dot{q})=\frac{1}{2}L_s\dot{q_L}^2-\frac{{{q}_C}^2}{2C};\\ 
\mathcal{D}(\dot{q},u)=\frac{1}{2}(\dot{q_L}^2R_s+(\dot{q}_L-\dot{q}_C)^2R).
\end{eqnarray}
For $u=0$ the expressions are: 
\begin{eqnarray}\label{a1}
\mathcal{T}(q,\dot{q}) = 0 ; \mathcal{V}(q) = \frac{{q_C}^2}{2C} ; \mathcal{F}(q,u) = \left[\begin{array}{cc}
0 & 0 
\end{array}\right]^T;\\\label{a2}
\mathcal{L}(q,\dot{q})=-\frac{{{q}_C}^2}{2C};\\ 
\mathcal{D}(\dot{q},u)=\frac{1}{2}{\dot{q}_C}^2R.\label{a3}
\end{eqnarray}
It is clear from the above equations that the EL parameters of the two circuits, generated by the two switch positions, do not result in identical values for magnetic and electric energies (amongst other quantities like $\mathcal{D}$ and $\mathcal{F}$). A combined set of EL equations from the two separate descriptions valid for both the switch positions can be written as follows: 
\begin{eqnarray} 
\mathcal{T}(q,\dot{q},u) = \frac{1}{2}L_s{(u\dot{q_L})}^2 ; \mathcal{V}(q) = \frac{{q_C}^2}{2C};\\ \label{aa}
\mathcal{F}(q,u) = \left[\begin{array}{cc}
uV_i & 0 
\end{array}\right]^T;\\ \label{ab}
\mathcal{L}(q,\dot{q},u)=\frac{1}{2}L_s{(u\dot{q_L})}^2-\frac{{{q}_C}^2}{2C};\\ \label{ac}
\mathcal{D}(\dot{q},u)=\frac{1}{2}({(u\dot{q_L})}^2R_s+(u\dot{q}_L-\dot{q}_C)^2R). 
\end{eqnarray}
It is evident from the above equations that $\mathcal{T}$ and $\mathcal{L}$ are functions of $u$ and this dependence is clearly shown. The set (\ref{aa}, \ref{ab}, \ref{ac}) when substituted in (\ref{elc},\ref{eld}) gives the following equations:
\begin{eqnarray}\label{x2}
u^2L_s\ddot{q}_L=-u^2\dot{q}_LR_s+(-u^2\dot{q}_L+u\dot{q}_C)R+uV_i,\\
\frac{q_C}{C}=(u\dot{q}_L-\dot{q}_C)R.\label{x3}
\end{eqnarray}
Substituting for $\dot{q}_C$ from (\ref{x3}) in (\ref{x2}) and rearranging gives the following switched state space model for the circuit in descriptor form:
\begin{eqnarray}\label{x4}
\begin{split}
\left[\begin{array}{cc}u&0\\ \\0&1\end{array}\right]
\left[\begin{array}{c}
\ddot{q_L}\\ \\
\frac{\dot{q}_C}{C}
\end{array}\right]=\left[\begin{array}{cc}-\frac{uR_s}{L_s}&\frac{-u}{L_s}\\ \\\frac{u}{C}&\frac{-1}{RC}\end{array}\right]\left[\begin{array}{c}
\dot{q_L}\\ \\
\frac{q_C}{C}\end{array}\right] \\ +\left[\begin{array}{c}
\frac{u}{L_s}\\ \\
0
\end{array}\right]V_i
\end{split}
\end{eqnarray}
Given the fact that $u$ can be either $0$ or $1$, (\ref{x4}) has been written by writing $u^2$ as $u$. Further, the descriptor form is indispensable because cancelling by $u$ on both sides of (\ref{x2}) is incorrect when $u=0$.

It is worth noting that although (\ref{x4}) does not reconstruct the second state equation $\dot{q}_L=0$ for $u=0$, it also does not give an erroneous representation like (\ref{xx}). The equation $\dot{q}_L=0$ for $u=0$ cannot be derived directly from the EL formulation because the relevant circuit representation (\ref{a1}-\ref{a3}) does not contain that information.

The above example shows that in switched circuits it is possible that the magnetic and electric energies of the system are also functions of the switching state, and ignoring this dependence while writing the EL equations for the circuit can lead to errors. The solution is to write the EL equations for each \textit{mode} of the circuit (corresponding to a particular value of $u$) and then write them as functions of $u$ for the overall circuit, as demonstrated above. In the next section we discuss the issue of incorporation of constraints while writing the EL equations for switched circuits.

\section{Incorporation of constraints in EL formulation}
\label{inco}
The use of constraint form of EL formulation is necessitated \citep{scherpen_2003} by the fact that the corresponding Lagrangian does not include complete information about the circuit. For example, in a circuit having more than one mesh, with current and charge of dynamic elements selected as generalised coordinates, the EL formulation represents a force balance, which means that the algebraic relation between various currents is missing from the formulation and has to be incorporated by adding constraints. In this section we examine this contention in detail and show that this is not always necessarily true. We demonstrate using examples that proper labelling of currents flowing in the circuit leads to automatic incorporation of Kirchoff's current law in the framework. Consequently, EL modelling in the unconstrained form is both possible as well as convenient.

The simplest case arises when there are $m$ constraints amongst the $n$ generalised coordinates (i.e., the charges and currents of dynamic elements $L$ and $C$). In this case it is optimum to choose the number of
generalized coordinates of the system as $N=n-m$. Subsequently, EL modelling can be done using the unconstrained form. This also ensures that a minimal representation for the system is directly obtained. A brief discussion of this case for EL modelling of three-phase pulse-width modulated AC-to-DC converters is given in \cite{tan}.

The second case arises when constraints are not between the generalised coordinates but stem from branch relations involving dissipative elements like resistances. In this case the constraints are automatically taken care of while writing the expression for the Rayleigh dissipation function $\mathcal{D}(\dot{q},u)$. An example of this case is the circuit studied in Section-\ref{inad} where the expression for $\mathcal{D}(\dot{q},u)$ in (\ref{diss}) is written by writing the current through the resistor $i_R$ in terms of the generalised current coordinates $\dot{q}_L$ and $\dot{q}_C$.

\begin{figure}[h!]
\vspace{-3mm}
\centering
\includegraphics[width=1\linewidth]{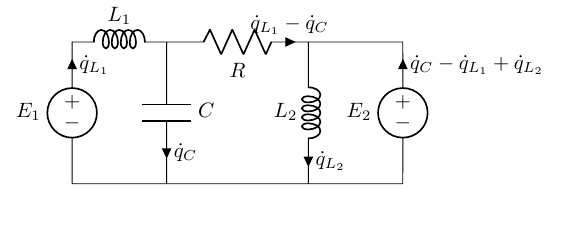}
\vspace{-10mm}
\caption{Example Circuit}  \label{figy}
\end{figure}


Another possibility is that the constraints stem from branch relations between currents through various voltage sources. In this case it is more convenient to write the branch current through the source as a linear combination of the generalised current coordinates and incorporate the forcing functions as part of electric field energy of the system. This is illustrated using the circuit shown in Fig. \ref{figy} for which following expression can be written for the energies and the dissipation function:
\begin{eqnarray}\label{r1}
\mathcal{T}(q,\dot{q}) = \frac{1}{2}(L_1\dot{q_{L_1}}^2+L_2\dot{q_{L_2}}^2);\\\label{r2} \mathcal{V}(q) = \frac{{q_{C}}^2}{2C}-E_1q_{L_1}-E_2(q_{C}-q_{L_1}+q_{L_2});\\
\mathcal{D}(\dot{q})=\frac{1}{2}R(\dot{q_{L_1}}-\dot{q_{C}})^2.
\end{eqnarray}
The labelling of currents ensures that all branch relations are taken care of and the unconstrained form of the EL equations can be directly applied. Further, as seen from (\ref{r2}), since the forcing function $E_2$ is not associated with any one generalised coordinate, it is more convenient to include forcing functions as part of $\mathcal{V}$ instead of writing a separate expression for $\mathcal{F}$.

\begin{figure}[h!]
\vspace{-3mm}
\centering
\includegraphics[width=0.58\linewidth, height = 0.4\linewidth ]{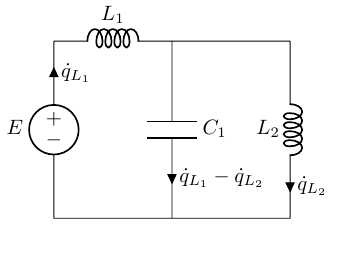}
\vspace{-10mm}
\caption{LC circuit}  \label{figx}
\end{figure}


Lastly, we examine one more case, where the constraints are not between the generalised coordinates and also cannot be incorporated using the dissipation function or electric field energy.  
The idea is demonstrated using the circuit shown in Fig. \ref{figx} taken from \cite{scherpen_2003}. The following equations can be written for the circuit:
\begin{eqnarray}\label{s1}
\mathcal{T}(q,\dot{q}) = \frac{1}{2}(L_1\dot{q_{L_1}}^2+L_2\dot{q_{L_2}}^2);\\\label{s2} \mathcal{V}(q) = \frac{(q_{L_1}-q_{L_2})^2}{2C_1}-Eq_{L_1}; \mathcal{D}(\dot{q})=0; \\
\label{s3} \begin{split}
\mathcal{L}(q,\dot{q})=\frac{1}{2}(L_1\dot{q_{L_1}}^2+L_2\dot{q_{L_2}}^2)\\-\frac{(q_{L_1}-q_{L_2})^2}{2C_1}+Eq_{L_1}.
\end{split}
\end{eqnarray}
Writing the equations like this automatically incorporates Kirchoff's current law and the constraint form is not required. Substitution of the equations (\ref{s1}-\ref{s3}) in the EL equations (\ref{Eq1}) alongwith the third state equation $\dot{q}_{C1}=\dot{q}_{L1}-\dot{q}_{L2}$ gives the state space representation for the circuit.

We have thus demonstrated that constraints can be incorporated using a standard procedure while writing the EL formulation for a circuit. It is best to label generalised current coordinates for the dynamic elements and then write all other currents as a linear combination of these generalised coordinates. This leads to automatic incorporation of Kirchoff's current law and reduces the amount of bookkeeping required for solving the problem. In the next section we describe modelling for power converters in more detail, and discuss their high-fidelity equivalent circuits. We then proceed to demonstrate the application of the EL formulation to these high fidelity models using the insights gained from the discussion in Sections \ref{inad} and \ref{inco}.
\section{High-fidelity equivalents of power converters and their EL modelling}
\subsection{Mathematical Modelling of power converters}
%

Mathematical models of systems are based on the understanding of their physical behaviour and are needed for their simulation and control. Basic circuit modelling of power electronic converters typically produces continuous-time, non-linear, time-varying models in the following form:

\begin{equation} \label{nl}
\dot{x}(t) = f_{\sigma(t)}(x(t),u(t))
\end{equation}
with the state $x \in R^n$, the input $u \in R^p$ and $\sigma(t) : [0,\inf) \rightarrow \Sigma$ is a right continuous function which is piecewise constant and which selects the index of the active system from the index set $\Sigma$ at each time instant $t$. The function $\sigma(t)$ is also called the mode-selector function. If the switching is only time-dependent, then corresponding to $\sigma(t)$ the following switching functions can be defined:
\begin{eqnarray}
q_i(t) : [0,\infty) \rightarrow \{0, 1\},\\
\Sigma_{i=1}^{m}q_i(t) = 1, \forall t \in [0,\infty).
\end{eqnarray}
Each index $\sigma(t)=i$ is called a mode and each mode defines a different dynamic behaviour of the system. Consequently, equation (\ref{nl}) can be re-written as:
\begin{equation}
\dot{x}(t) = \Sigma_{i=1}^{m}q_i(t)f_i(x(t),u(t)).
\end{equation}
Depending on the application, sufficiently accurate assumptions not affecting the validity of the models can be made. Switches can be considered \textit{ideal}, consequently modelling them as resistances of zero and infinity during turn-on and turn-off, respectively. The switching time can be considered to be infinitely short. Generators can be considered \textit{ideal} and passive circuit elements (R,L and C) can be considered linear and time-invariant.

Applying these assumptions would lead to a switched state-space linear time invariant (LTI) model of the following form:
\begin{equation}
\dot{x}(t) = A_{\sigma(t)}x(t)+B_{\sigma(t)}u(t).
\end{equation}
Using the switching function notation, this can be re-written as:
\begin{equation}
\dot{x}(t) = \Sigma_{i=1}^{m}q_i(t)(A_ix(t)+B_iu(t)).
\end{equation}
However, if the switching is state dependent as well (this happens when devices like diodes are present), then the switching function would be written as $q_i(t, x(t))$, and the model ceases to remain linear. A further complication in the model
is introduced by the fact that a pulse width modulated (PWM) converter switches in response to a modulating signal $m(t)$. Augmenting the model to represent the relation between $q_i(t)$
and $m(t)$ introduces additional non-linearity and time varying behaviour.

In addition to switched state-space models, other modelling approaches include ones based on circuit averaging, sampled data and dynamic phasors. For details see \citep{baccha, maksim}.

\subsection{High-fidelity equivalents of power converters}
EL modelling of power converters has been confined so far to idealized equivalent circuits, especially idealized equivalents for switching devices and dynamic elements. Parasitics are not included, and hence the models obtained are not suitable for component-level study to examine phenomena such as device voltage and current transients. In this section, we extend EL modelling of power electronic controllers to complex, non-ideal, high-fidelity (H-F) descriptions of these converters. These H-F state-space models are otherwise difficult to obtain using classical methods.

\begin{figure}[h!]
\vspace{-10mm}
\centering
\includegraphics[width=0.65\linewidth]{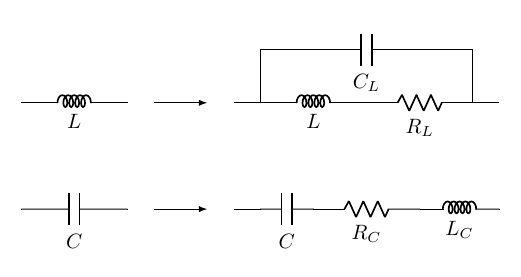}
\vspace{-5mm}
\caption{H-F inductor and capacitor model}  \label{fig41}
\end{figure}

H-F models of power converters \citep{multiresolution, hkhan-hifi} are obtained when stray inductances and capacitances of passive components are also considered. An inductor is replaced \citep{selfCofL} by $R_{L}$, $C_{L}$  and the bulk inductance $L$, and the capacitor is modelled with  an  equivalent  series resistance $R_{c}$ and  inductance $L_{c}$  in addition to bulk capacitance $C$, shown in Fig. \ref{fig41}. The Diode is modelled as a piece-wise linear model. Its voltage drop $V_{D}(u)$ is a function of the switch state $u$ of the diode, $V_{D}(0)= 0$, $V_{D}(1)= V_{d_{on}}$, where $u=0$ for \textit{off} state and $u=1$ for \textit{on} state of the diode. $R_{d}(u)$ is used to model the linear portion of the diode characteristic curve, This resistance has two values based on the diode state: $R_{d}(1)$ = $R_{d_{\emph{on}}}$, $R_{d}(0)$ = $R_{d_{\emph{off}}}$. The junction capacitance  $C_{d}$ models reverse recovery effects.  

\begin{figure}[h]
\vspace{-5mm}
\centering
\includegraphics[width=0.65\linewidth]{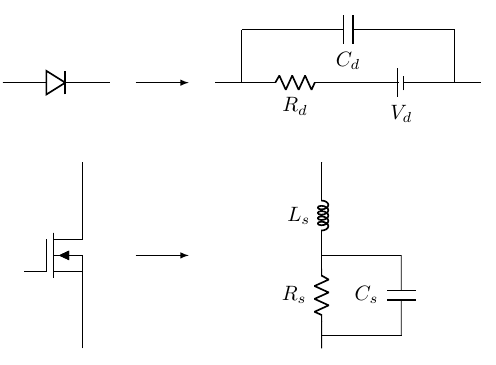}
\caption{H-F model of Diode and MOSFET}  \label{fig42}
\end{figure}

The MOSFET drain to source characteristics is modeled by $R_{s}(u_{m})$, $L_{s}$ and $C_{s}$, with $R_{s}(u_{m})$ being the switch mode dependent resistance, where $u_{m}$ is the switch state of MOSFET ($u_{m}=0$ for \textit{off}, $u_{m}=1$ for \textit{on}),  see Fig. \ref{fig42}.

\subsection{EL modelling of H-F equivalents}
\subsubsection{H-F model of diode rectifier}
\begin{figure}[h!] 
\centering
\includegraphics[width=9cm,height=6cm]{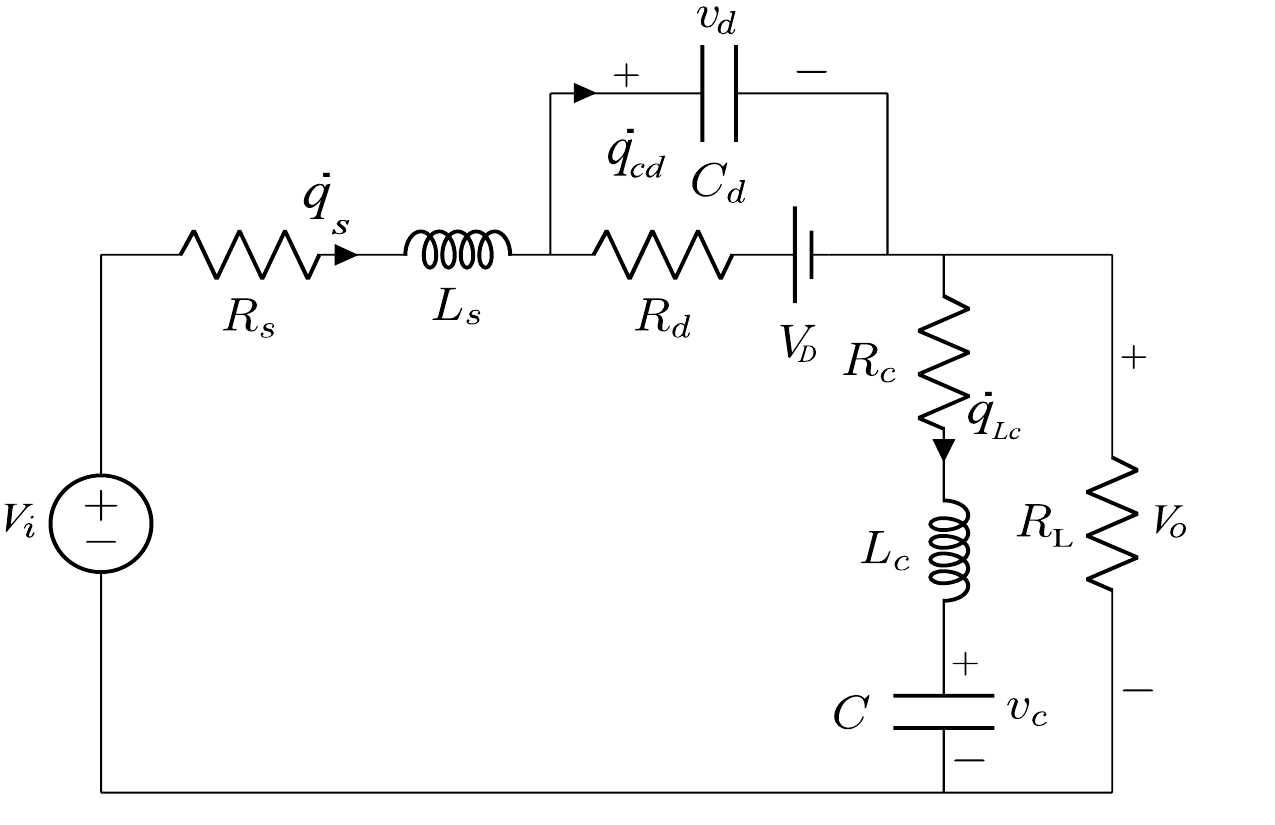}
\caption{H-F model of diode rectifier} \label{fig2}
\end{figure}

The H-F equivalent of the simple diode rectifier circuit shown in Fig. \ref{fig0} is shown in Fig. \ref{fig2}. 
It is clear from Fig. \ref{fig2} that only currents in the dynamic elements are labelled, all other currents would be written as a linear combination of these currents. The generalised coordinate set includes these currents and corresponding charges, i.e., the set $(q_s, q_{Lc}, q_{cd}, \dot{q}_s, \dot{q}_{Lc}, \dot{q}_{cd})$, and the following expressions can be written for $\mathcal{V}(q,u),\mathcal{T}(q,\dot{q}),\mathcal{L}(q,\dot{q},u)$ and $\mathcal{D}(\dot{q},u)$.
\begin{eqnarray}   \label{ed}
 \mathcal{T}(q,\dot{q})=\frac{1}{2} L_{s}\dot{q_{s}}^{2}+\frac{1}{2} L_{c}\dot{q}_{Lc}^{2},\\ \label{ee}
 \begin{split}
 \mathcal{V}(q,u)=\frac{1}{2C_{d}} q_{cd}^{2}+ \frac{1}{2C} q_{Lc}^{2}-V_iq_{s}\\+V_{D}(u)(q_{s}-q_{cd}),
 \end{split}
 \end{eqnarray}

 \begin{multline} \label{ef}
 \mathcal{L}(q,\dot{q},u)=\frac{1}{2} L_{s}\dot{q_{s}}^{2}+\frac{1}{2} L_{c}\dot{q}_{Lc}^{2} -\frac{1}{2C_{d}} q_{cd}^{2}-\frac{1}{2C} q_{Lc}^{2} \\
 + V_iq_{s}-V_{D}(u)(q_{s}-q_{cd}),
 \end{multline}

 \begin{multline}  \label{eg}
\mathcal{D}(\dot{q},u)=\frac{1}{2} R_{s}\left(\dot{q_{s}}\right)^{2}+\frac{1}{2} R_{c}\left(\dot{q}_{Lc}\right)^{2}+\frac{1}{2} R_{d}(u)\left(\dot{q_{s}}-\dot{q}_{cd}\right)^{2} \\
+\frac{1}{2} R_{L}\left(\dot{q_{s}}-\dot{q}_{Lc}\right)^{2}.
\end{multline}
It maybe noted that in the above equations the voltage drop $V_D$ and resistance $R_d$ of the diode are written as $V_D(u)$  and $R_d(u)$, respectively, to clearly illustrate their dependence on the switching function $u$. Consequently, this dependence is reflected in the expressions for $\mathcal{V}$, $\mathcal{L}$ and $\mathcal{D}$ given above. Further, current constraints have been incorporated automatically in these equations, so $\mathcal{A}(\dot{q},u)=0$.

Using equations (\ref{ed}, \ref{ee}, \ref{ef}, \ref{eg}) in equation (\ref{elc}) and solving for $q=q_{s}$ we get:

  $$
 \ddot{q_{s}}=\frac{(-R_{s}-R_{L})}{L_{s}}\dot{q_{s}}-\frac{q_{cd}}{L_sC_d}+\frac{R_{L}}{L_{s}}\dot{q}_{Lc}+\frac{V_{i}}{L_{s}},
  $$
  for $q=q_{Lc}$ we obtain
  $$
 \ddot{q}_{Lc}=\frac{R_{L}}{L_{c}}\dot{q}_{s}-\frac{(-R_{L}-R_{c})}{L_{c}}\dot{q}_{Lc}-\frac{q_{Lc}}{CL_{s}},
  $$
  while for $q=q_{cd}$ we obtain
  $$
 \dot{q}_{cd}=\dot{q}_{s}-\frac{q_{cd}}{R_{d}(u)C_{d}}+\frac{V_{D}(u)}{R_{d}(u)}.
  $$
  Selecting the state vector as $\mathbf{x}=\left( i,v_{d},i_{Lc},v_{c} \right)^{T}= \left( \dot{q_{s}},\frac{{q}_{cd}}{C_{d}},\dot{q}_{Lc},\frac{q_{Lc}}{C} \right)^{T}$, writing $V_D(u)$ as $uV_{d_{on}}$ and re-writing the input vector $u$ as $u=\left( V_{i},V_{d_{on}}\right)^{T}$  the final state-space model obtained is:

  $$
\mathbf{A(u)}=\begin{bmatrix}
\frac{\left(-R_{s}-R_{L}\right)}{L_{s}} & \frac{-1}{L_{s}} &\frac{R_{c}}{L_{s}}&0\\ \\ \frac{1}{C_{d}} & -\frac{1}{R_{d}(u)C_{d}} &0&0 \\ \\
\frac{R_{L}}{L_{c}}&0 &\frac{\left(-R_{L}-R_{c}\right)}{L_{c}}&\frac{-1}{L_{c}}\\ \\ 0&0 &\frac{1}{C}&0\end{bmatrix} ;
$$  
$$
\mathbf{B(u)}=\begin{bmatrix}
\frac{1}{L_{s}}&0\\ \\  0&\frac{u}{R_{d}(u)C_{d}} \\ \\  0&0
\\ \\  0&0\end{bmatrix};\mathbf{C}=\left[I\right]_{4\times 4} ;\mathbf{D}=\left[0\right]_{4\times 2}.
$$ 
\subsubsection{H-F model of DC-DC boost converter}
A DC-DC converter is shown in Fig. \ref{fig3}. Replacing the switches and the dynamic elements by their H-F equivalents gives the circuit configuration shown in Fig. \ref{fig4}. As before, the generalised coordinates selected are currents and charges of dynamic elements with the currents shown in Fig. \ref{fig4}. Writing the equations like we did for the previous example, we obtain:
\begin{figure}[h!] 
\centering
\includegraphics[width=0.7\linewidth]{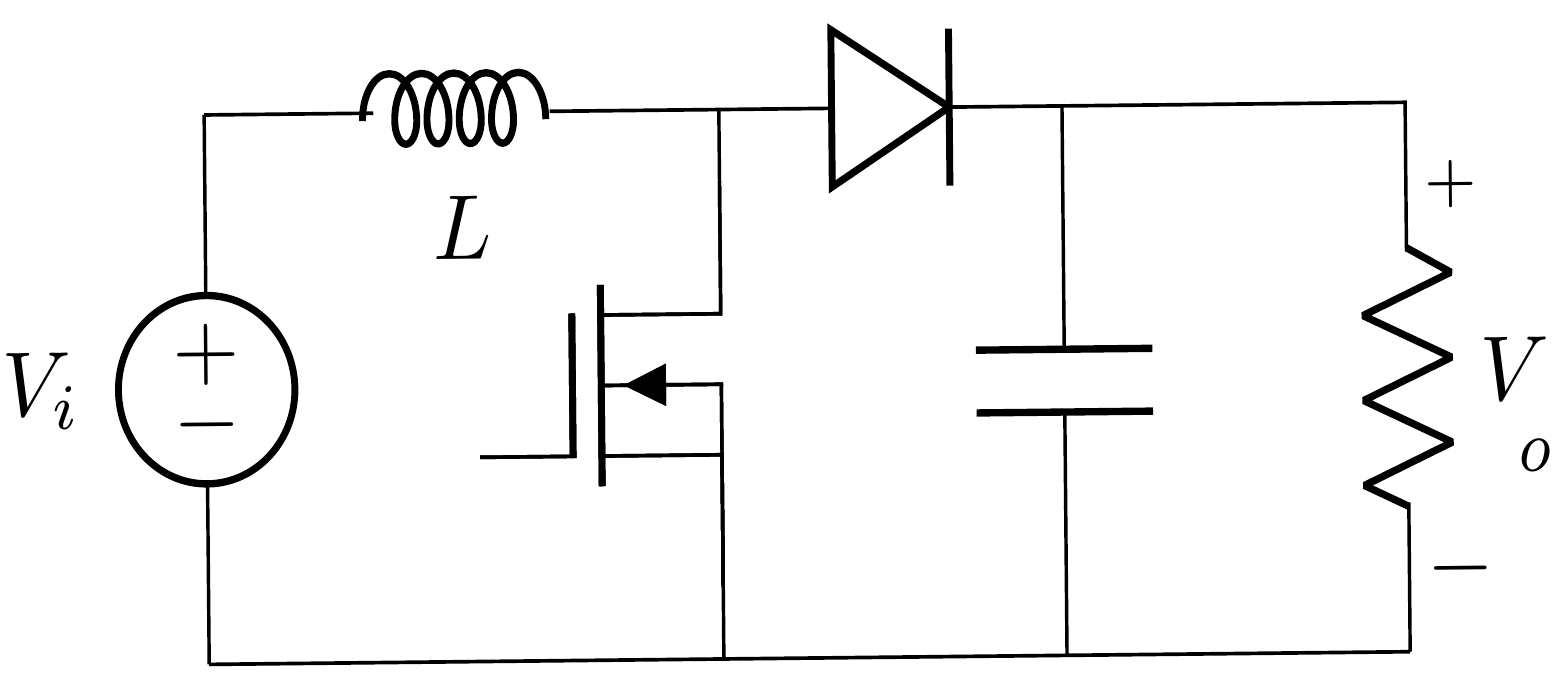} 
\caption{DC-DC Boost-converter} \label{fig3}
\end{figure}

\begin{equation}  \label{l7}
 \mathcal{T}(q,\dot{q})=\frac{1}{2} L\dot{q_{1}}^{2}+\frac{1}{2} L_{s}\dot{q_{2}}^{2} +\frac{1}{2} L_{c}\dot{q_{3}}^{2},
 \end{equation}
 \begin{eqnarray} \label{l8}
 \begin{split}
 \mathcal{V}(q,u_d)=\frac{1}{2C} q_{3}^{2}+\frac{1}{2C_{s}} q_{4}^{2}+\frac{1}{2C_{d}} q_{5}^{2}-V_iq_{1}\\+V_{D}(u_{d})(q_{1}-q_{2}-q_{5}),
 \end{split}
 \end{eqnarray}
 \begin{multline}  \label{l9}
 \mathcal{L}(q,\dot{q},u_d)=\frac{1}{2} L\dot{q_{1}}^{2}+\frac{1}{2} L_{s}\dot{q_{2}}^{2} +\frac{1}{2} L_{c}\dot{q_{3}}^{2}-\frac{1}{2C} q_{3}^{2}-\frac{1}{2C_{s}} q_{4}^{2} \\ -\frac{1}{2C_{d}} q_{5}^{2}+V_iq_{1}-V_{D}(u_{d})(q_{1}-q_{2}-q_{5}),
 \end{multline}
 \begin{multline}  \label{mm}
\mathcal{D}(\dot{q},u_d)=\frac{1}{2} R_{L}\left(\dot{q_{1}}\right)^{2}+\frac{1}{2} R_{s}(u_{m})\left(\dot{q_{2}}-\dot{q_{4}}\right)^{2}+\frac{1}{2} R_{c}\left(\dot{q_{3}}\right)^{2} \\+\frac{1}{2} R_{d}(u_{d})\left(\dot{q_{1}}-\dot{q_{2}}-\dot{q_{5}}\right)^{2}+\frac{1}{2} R_{o}\left(\dot{q_{1}}-\dot{q_{2}}-\dot{q_{3}}\right)^{2}.
\end{multline}

\begin{figure}[ht] 
\vspace{-20mm}
\centering
\includegraphics[width=0.9\linewidth]{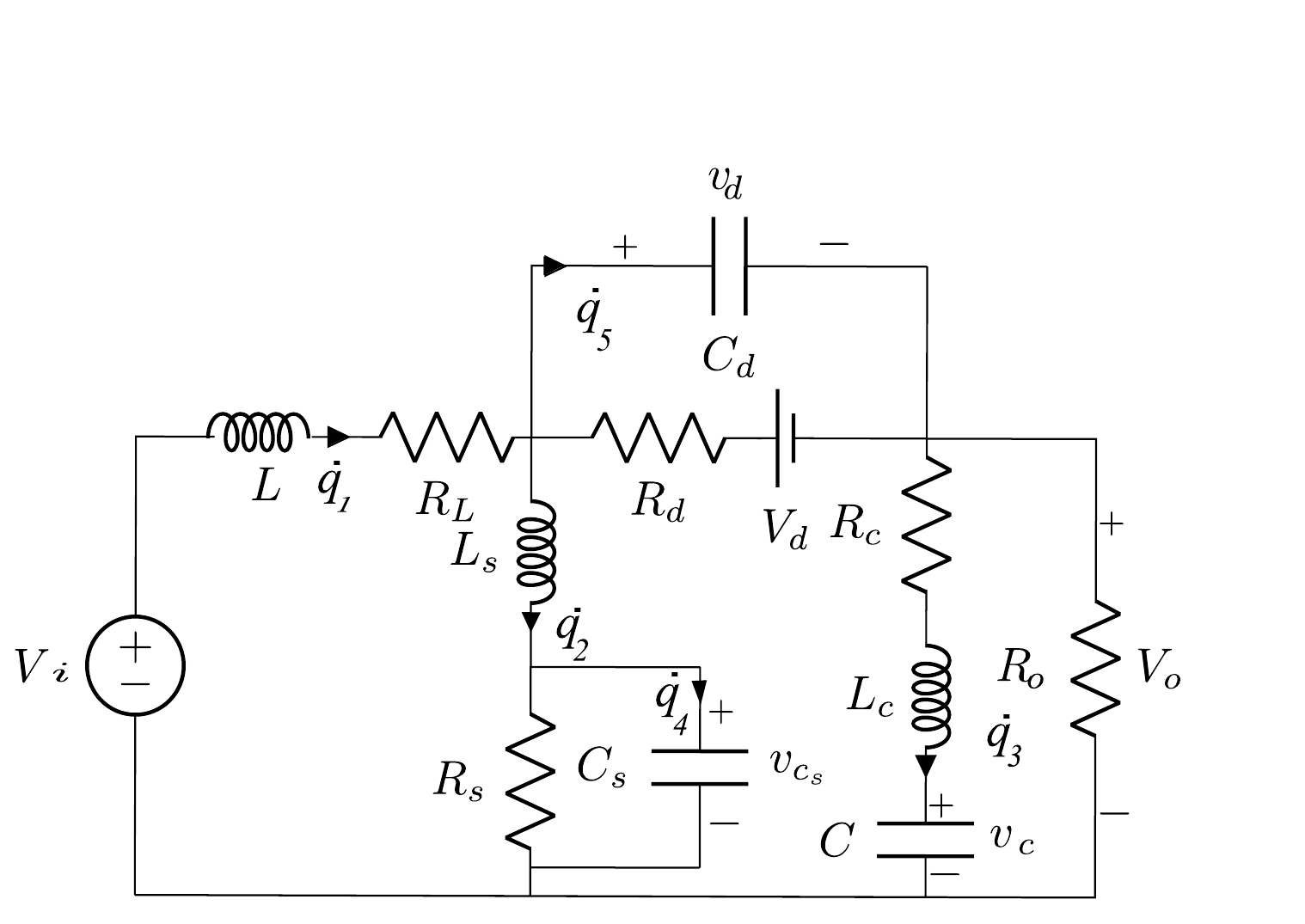}
\caption{H-F model of DC-DC Boost converter} \label{fig4} 
\end{figure}
Again, in the above equations, $V_D(u_{d}), R_d(u_{d})$ illustrate the dependence of the diode parameters on the diode switching function $u_{d}$ and $R_{s}(u_{m})$ illustrates the dependence of the MOSFET resistance on the MOSFET switching function $u_{m}$. Putting equations (\ref{l7}-\ref{mm}) in  (\ref{elc}) and solving for $q=q_{1}$ we get:
  \begin{equation*}
 \ddot{q_{1}}=\frac{(-R_{L}-R_{o})}{L}\dot{q_{1}}+\frac{R_{o}}{L}\dot{q_{2}}+\frac{R_{o}}{L}\dot{q_{3}}-\frac{q_5}{LC_d}+\frac{V_{i}}{L},
 \end{equation*}
  for $q=q_{2}$ we obtain:
  \begin{equation*}
 \ddot{q_{2}}=\frac{R_{o}}{L_{s}}\dot{q_{1}}-\frac{R_{o}}{L_{s}}\dot{q_{2}}-\frac{R_{o}}{L_{s}}\dot{q_{3}}-\frac{q_4}{L_sC_s}+\frac{q_5}{L_sC_d},
  \end{equation*}
 for $q=q_{3}$ we obtain:
  \begin{equation*}
 \ddot{q_{3}}=\frac{R_{o}}{L_{c}}\dot{q_{1}}-\frac{R_{o}}{L_{c}}\dot{q_{2}}-\frac{q_{3}}{CL_c}+\frac{\left(-R_{c}-R_{o}\right)}{L_{c}}\dot{q_{3}},
  \end{equation*}
  for $q=q_{4}$, we obtain:
  \begin{equation*}
 \dot{q_{4}}=\dot{q_{2}}-\frac{q_{4}}{R_{s}(u_{m})C_{s}},
 \end{equation*}
  and for $q=q_{5}$ we obtain:
  \begin{equation*}
  \dot{q_5}=\dot{q_1}-\dot{q_2}-\frac{q_{5}}{R_{d}(u_{d})C_d}+\frac{V_{D}(u_{d})}{R_{d}(u_{d})}.
 \end{equation*}
Selecting the state vector as $\mathbf{x}=\left( i,v_{c},i_{Ls},v_{cs},i_{Lc},v_{d} \right)^{T}= \left( \dot{q_{1}},\frac{q_{3}}{C},\dot{q_{2}},\frac{q_{4}}{C_{s}},\dot{q_{3}},\frac{q_{5}}{C_{d}} \right)^{T}$, re-writing $V_D(u_{d})$ as $u_{d}V_{d_{on}}$ and writing the input vector $u$ as $u=(V_{i},V_{d_{on}})^{T}$, we obtain the final state-space model (see Appendix). It is worth noting that $u_{m}$ and $u_{d}$ are complementary to each other, so the state matrix derived can be written as a function of $u_{d}$ alone.
\section{Simulation results}
In this section, we present simulation results for two high-fidelity power converters, taking into account the non-idealities of circuit components. We also provide model parameters and a brief discussion of the results. Additionally, for reproducibility, we include the code at the following repository: \href{https://github.com/ShakirSofi/H-F-Dc-Dc-Power-converters}{https://github.com/ShakirSofi/H-F-Dc-Dc-Power-converters}. 
\subsection{H-F model diode rectifier} 
\begin{figure}[h]
\includegraphics[width=\linewidth]{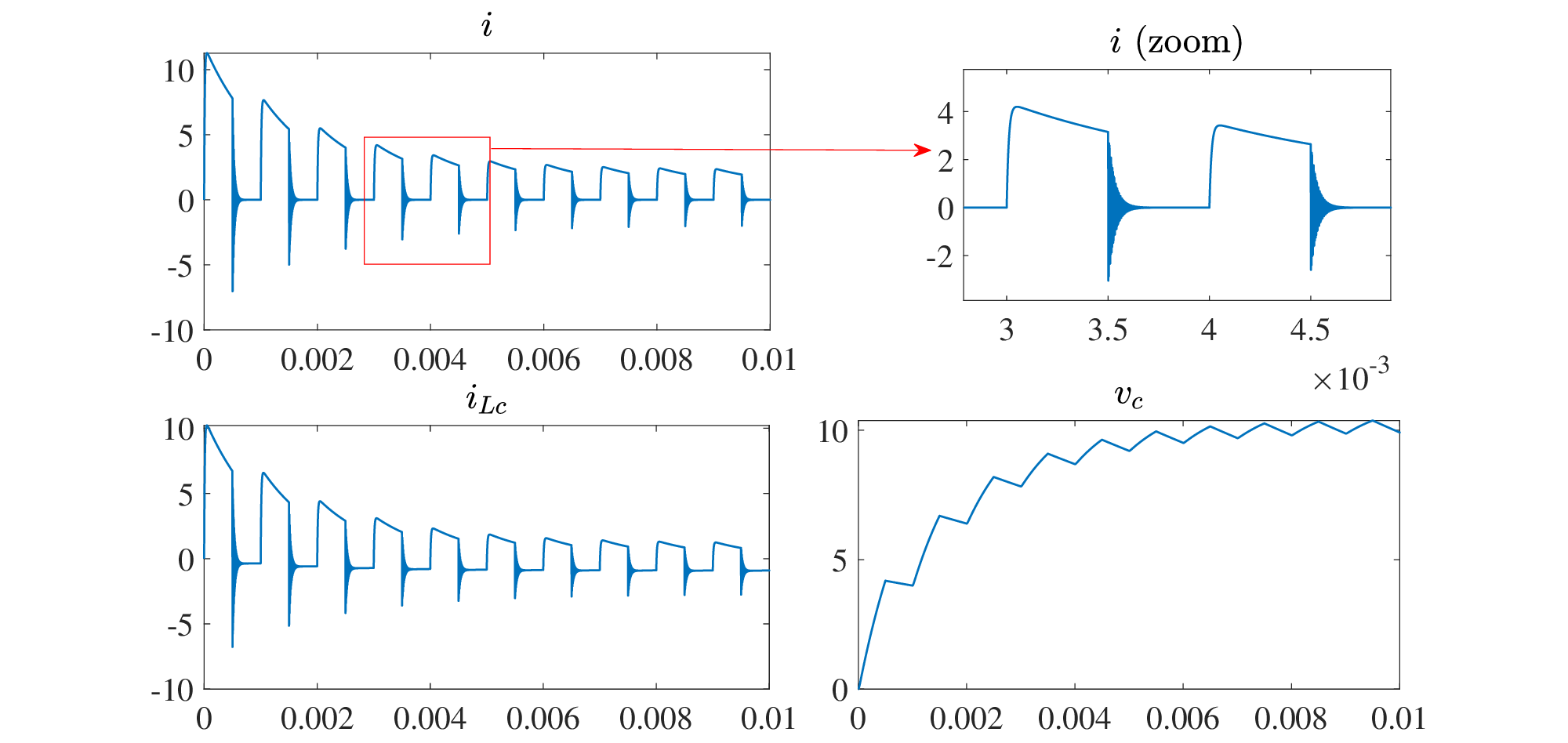} 
\caption{Simulation results of H-F model of diode rectifier} \label{Diodetsim}
\end{figure}

\paragraph{Parameters}
The input is a square wave signal of $\pm 12$ V with a switching frequency of $1$ kHz, and the other parameters are set as follows: $R_{s} = 0.01 \, \Omega$, $C = 1 \, \text{mF}$, $L_{c} = 10 \, \mu\text{H}$, $R_{L} = 10 \, \Omega$; $L_{s} = 10 \, \mu\text{H}$, $R_{d_{\text{off}}} = 10k \, \Omega$, $R_{d_{\text{on}}} = 0.05 \, \Omega$, $C_{d} = 10 \, \text{nF}$, $R_{c} = 1 \, \Omega$, $R_{L} = 10 \, \Omega$, and $L_{c} = 10 \, \text{nH}$.
\paragraph{Results}
Figure~\ref{Diodetsim} shows that the inductor current $i$ changes smoothly when the switch goes from the OFF to the ON state and starts storing energy. The current increases as the magnetic field collapses when the switch is turned OFF. The inductor will release this energy by increasing voltage. On the right, the figure shows damped oscillations, and the peak oscillation occurs at the start when the switch is turned OFF, i.e., the current decreases significantly, causing an increase in voltage. Similarly, the bottom left plot shows the voltage buildup across the capacitor. 

\subsection{H-F DC-DC boost converter} 
\paragraph{Parameters} The input voltage is $V_{i} = 10$ V, and the parameters are as follows: $R_L = 0.1 \, \Omega$, $C = 42 \, \mu \text{F}$, $L = 1.6 \, \text{mH}$, $L_{s} = 20 \, \text{nH}$, $R_{s_{\text{off}}} = 2 \, M\Omega$, $R_{s_{\text{on}}} = 0.2 \, \Omega$, $R_{d_{\text{on}}} = 50 \, m\Omega$, $C_{s} = 200 \, pF$, $C_{d_{\text{on}}} = 15 \, mF$, $R_{c} = 0.4 \, \Omega$, $R_{d_{\text{off}}} = 40 \, M\Omega$, $L_{c} = 100 \, pH$, $d(t) = 0.5$, and switching frequency $f = 50$ kHz.

\begin{figure}[h!]
\includegraphics[width=\linewidth]{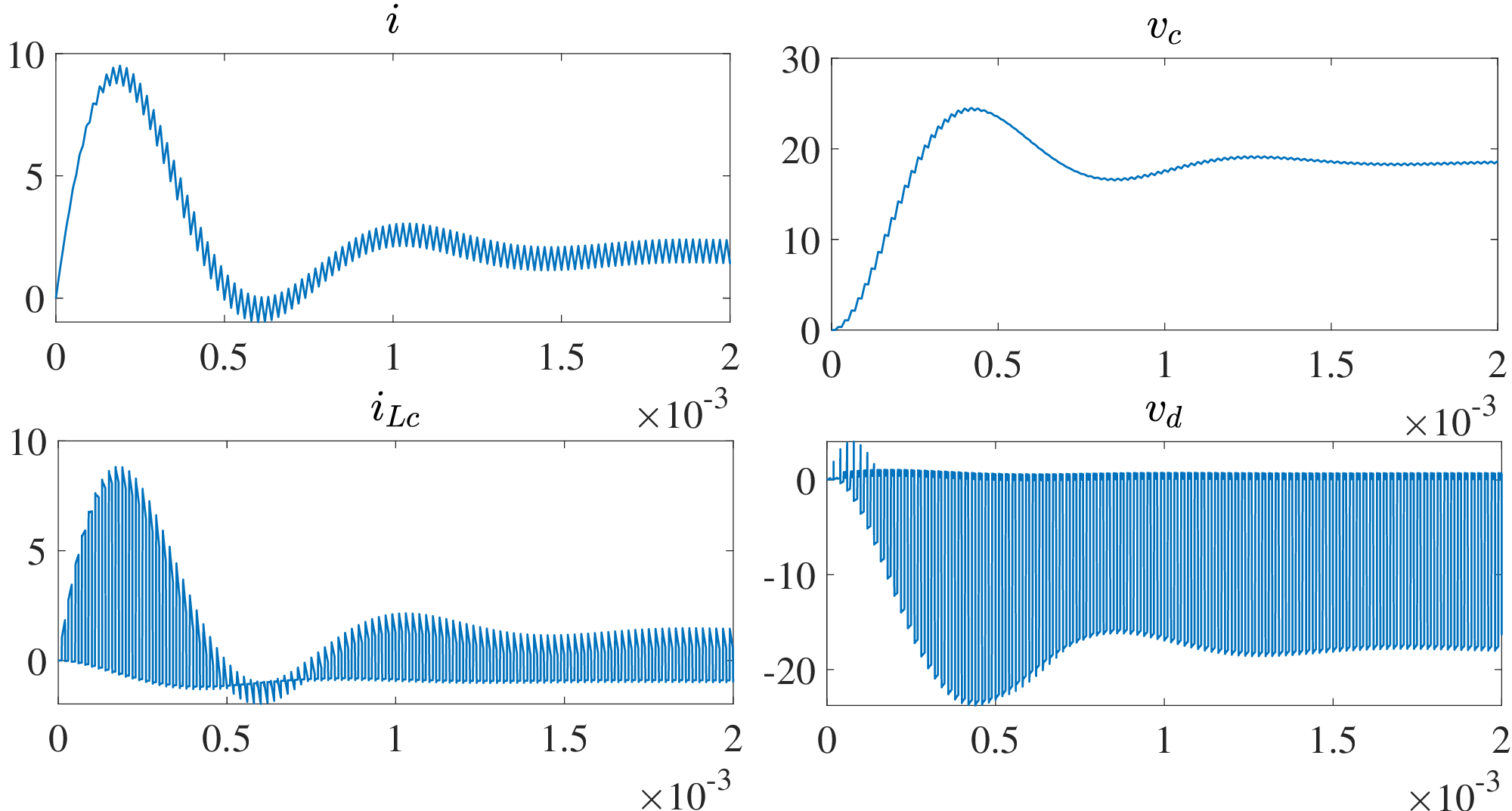} 
\caption{Simulation results of H-F model of DC-DC Boost Converter}\label{Boostsim}
\end{figure}

\paragraph{Results} In the ideal scenario, the output voltage at a duty ratio of 0.5 should be $20V$ ($= \frac{V_{in}}{1-d}$). However, due to non-idealities, the measured voltage across $R_o$ is $18.22V$ in the steady state, and the steady-state inductor current obtained is $2.13A$, as shown in Figure~\ref{Boostsim}. The plots in the bottom row show the current across inductor $ I_{Lc}$ and the voltage across diode $v_d$. It can be inferred that the steady-state diode voltage is approximately  $-v_c$ when the switch is ON, 0.7 (diode bias voltage) otherwise. This can be seen in the plots and impulsive behaviour as the switching frequency is high.

\section{Conclusions}
We have revisited the procedure to be followed for Lagrangian modelling of switching circuits and pointed out errors in the erstwhile assumption that energies and consequently the Lagrangian cannot depend on the switching function. We have also demonstrated how to write down the EL formulation correctly without resorting to this assumption and obtain the correct state-space representation. We have examined the issue of incorporation of constraints in the EL formulation and shown how it can be done by an appropriate labelling of circuit currents in terms of generalised current coordinates. Lastly, with the help of the above-mentioned insights, we have extended Lagrangian modelling to high-fidelity equivalent circuits of power electronic converters. It is shown that despite the complex nature of these circuits, state-space models can be derived in an easy and systematic manner using the EL framework. Additionally, convincing numerical simulations are also presented to capture the non-ideal behaviour of high-fidelity circuits.  

\section*{Declarations}
\subsection*{Funding and Competing interests} This research received no external funding. The authors declare that they have no competing interests relevant to the article.

\appendix
\section*{Appendix} 
\begin{equation*}
\mathbf{A(u_{d}, u_{m})}=\begin{bmatrix}
\frac{\left(-R_{L}-R_{o}\right)}{L}&0&\frac{R_{o}}{L}&0&\frac{R_{o}}{L}&\frac{-1}{L}\\ \\  0&0&0&0&\frac{1}{C} &0 \\ \\  \frac{R_{o}}{L_{s}}&0& -\frac{R_{o}}{L_{s}}&\frac{-1}{L_{s}}&-\frac{R_{o}}{L_{s}}&\frac{1}{L_{s}}
\\ \\  0&0&\frac{1}{C_{s}}&-\frac{1}{R_{s}(u_{m})C_{s}} &0 &0 \\ \\ 
\frac{R_{o}}{L_{c}}&-\frac{1}{L_{c}}&-\frac{R_{o}}{L_{c}}&0&\frac{\left(-R_{c}-R_{o}\right)}{L_{c}}&0 \\ \\
\frac{1}{C_{d}}&0&-\frac{1}{C_{d}}&0&0&-\frac{1}{R_{d}(u_{d})C_{d}}
\end{bmatrix};
\end{equation*}

\begin{equation*}
\mathbf{B(u_{d})}=\begin{bmatrix}
\frac{1}{L}&0\\ \\  0&0 \\ \\  0&0
\\ \\  0&0\\ \\ 
0&0 \\ \\ 0 & \frac{u_{d}}{R_{d}(u_{d})C_{d}}
\end{bmatrix};
\mathbf{C}=\left[I\right]_{6\times 6} ;\mathbf{D}=\left[0\right]_{6\times 2}.
\end{equation*}


 \bibliographystyle{elsarticle-num} 
 \bibliography{main}

\begin{thebibliography}{10}
\expandafter\ifx\csname url\endcsname\relax
  \def\url#1{\texttt{#1}}\fi
\expandafter\ifx\csname urlprefix\endcsname\relax\def\urlprefix{URL }\fi
\expandafter\ifx\csname href\endcsname\relax
  \def\href#1#2{#2} \def\path#1{#1}\fi

\bibitem{mohan}
N.~{Mohan}, W.~P. {Robbins}, T.~M. {Undeland}, R.~{Nilssen}, O.~{Mo},
  Simulation of power electronic and motion control systems-an overview,
  Proceedings of the IEEE 82~(8) (1994) 1287--1302.

\bibitem{tan}
G.~Tan, H.~Chen, X.~Zhang, Comments on “lagrangian modeling and
  passivity-based control of three-phase ac/dc voltage-source converters”,
  IEEE Transactions on Industrial Electronics 55~(4) (2008) 1881--1882.

\bibitem{bloch}
A.~{Bloch}, Electromechanical analogies and their use for the analysis of
  mechanical and electromechanical systems, Journal of the Institution of
  Electrical Engineers - Part I: General 92~(52) (1945) 157--169.

\bibitem{pbm}
D.~Jeltsema, J.~M. Scherpen, Dynamics and Control of Switched Electronic
  Systems, Springer-Verlag, London, 2012, Ch. Power based modelling.

\bibitem{ortega_automatica}
G.~Escobar, A.~J. van~der Schaft, R.~Ortega, A hamiltonian viewpoint in the
  modeling of switching power converters, Automatica 35~(3) (1999) 445 -- 452.

\bibitem{Cuk}
R.~D. {Middlebrook}, S.~{Cuk}, A general unified approach to modelling
  switching-converter power stages, in: 1976 IEEE Power Electronics Specialists
  Conference, 1976, pp. 18--34.

\bibitem{scherpen_2003}
J.~M. Scherpen, D.~Jeltsema, J.~Klaassens, Lagrangian modeling of switching
  electrical networks, Systems and Control Letters 48~(5) (2003) 365 -- 374.

\bibitem{scherpen_2019}
D.~Jeltsema, J.~M. Scherpen, On the existence of lagrangians for clarke and
  park transformed switched-mode electrical networks, IFAC-PapersOnLine 52~(16)
  (2019) 90--95, 11th IFAC Symposium on Nonlinear Control Systems NOLCOS 2019.

\bibitem{umetani2016}
K.~Umetani, Lagrangian method for deriving electrically dual power converters
  applicable to nonplanar circuit topologies, IEEJ Transactions on Electrical
  and Electronic Engineering 11~(4) (2016) 521--530.

\bibitem{russer2012}
J.~A. Russer, P.~Russer, Lagrangian and hamiltonian formulations for classical
  and quantum circuits, IFAC Proceedings Volumes 45~(2) (2012) 439--444.

\bibitem{ortega_book}
R.~Ortega, J.~A.~L. Perez, P.~J. Nicklasson, H.~Sira-Ramirez, Passivity-based
  Control of Euler-Lagrange Systems, Springer-Verlag, London, 1998.

\bibitem{scherpen_99}
J.~Scherpen, J.~Klaassens, L.~Ballini, Lagrangian modeling and control of
  dc-to-dc converters, in: Proceedings of the INTELEC'99, University of
  Groningen, Research Institute of Technology and Management, Groningen, 1999.

\bibitem{yildiz2009}
H.~A. Yildiz, L.~Goren-Sumer, Lagrangian modeling of dc-dc buck-boost and
  flyback converters, in: 2009 European Conference on Circuit Theory and
  Design, 2009, pp. 245--248.

\bibitem{jmeisel}
J.~Meisel, Principles Of Electromechnical-energy Conversion, Krieger Publishing
  Company, Florida, 1984.

\bibitem{baccha}
S.~Bacha, I.~Munteanu, A.~I. Bratcu, Power Electronic Converters Modeling and
  Control, Advanced Textbooks in Control and Signal Processing,
  Springer-Verlag, London, 2014.

\bibitem{maksim}
D.~{Maksimovic}, A.~M. {Stankovic}, V.~J. {Thottuvelil}, G.~C. {Verghese},
  Modeling and simulation of power electronic converters, Vol.~89, 2001, pp.
  898--912.

\bibitem{multiresolution}
P.~L. {Chapman}, Multi-resolution switched system modeling, in: 2004 IEEE
  Workshop on Computers in Power Electronics, 2004. Proceedings., 2004, pp.
  167--172.

\bibitem{hkhan-hifi}
H.~{Khan}, M.~A. {Bazaz}, S.~A. {Nahvi}, Model order reduction of power
  electronic circuits, in: 2017 6th International Conference on Computer
  Applications In Electrical Engineering-Recent Advances (CERA), 2017, pp.
  450--455.

\bibitem{selfCofL}
A.~{Massarini}, M.~K. {Kazimierczuk}, Self-capacitance of inductors, IEEE
  Transactions on Power Electronics 12~(4) (1997) 671--676.

\end{thebibliography}

\end{document}